\documentclass[onecolumn,notitlepage,aps,pra,superscriptaddress,11pt,showpacs]{revtex4-2}
\usepackage[usenames,dvipsnames]{color}
\usepackage[english]{babel}
\usepackage{epsfig}
\usepackage{graphicx,subfigure}
\usepackage{float}
\usepackage{tikz}
\usepackage{pgfplots}
\pgfplotsset{compat=1.12}
\usetikzlibrary{calc}
\usetikzlibrary{decorations.markings}
\usetikzlibrary{decorations.pathmorphing}
\usetikzlibrary{positioning,shapes,arrows,decorations.markings,calc}
\usepackage{longtable}
\usepackage{multirow}
\usepackage{arydshln}
\usepackage{hyperref}
\usepackage{amsmath,bbm,bm,amssymb}
\usepackage[mathcal,mathscr]{eucal}
\DeclareFontFamily{OT1}{pzc}{}
\DeclareFontShape{OT1}{pzc}{m}{it}{<-> s * [1.10] pzcmi7t}{}
\DeclareMathAlphabet{\mathpzc}{OT1}{pzc}{m}{it}
\usepackage[latin1]{inputenc}
\usepackage[T1]{fontenc}
\usepackage{times}
\usepackage{lmodern}
\usepackage[Euler]{upgreek}
\usepackage{enumitem}
\setenumerate[1]{itemsep=0pt,partopsep=0pt,parsep=0pt,topsep=0pt}
\setitemize[1]{itemsep=0pt,partopsep=0pt,parsep=0pt,topsep=0pt}
\setdescription[1]{itemsep=0pt,partopsep=0pt,parsep=0pt,topsep=0pt}

\usepackage{bookmark}

\usepackage{textcomp}

\setcitestyle{numbers,square,citesep={,\kern-.24em}}

\hypersetup{
  colorlinks=false,
  pdfborder={0 0 0},
}

\tikzset{midarrow/.style={decoration={
  markings,
  mark=at position #1 with {\arrow{angle 45}}},postaction={decorate}}
}
\tikzset{waved/.style={decorate,decoration=snake}}

	\definecolor{aaltoBlack}{RGB}{0,0,0}%
	\definecolor{aaltoGray}{RGB}{146,139,129}%
	\definecolor{aaltoRed}{RGB}{237,41,57}%
	\definecolor{aaltoBlue}{RGB}{0,101,189}%
	\definecolor{aaltoYellow}{RGB}{254,203,0}%
	\definecolor{aaltoPurple}{RGB}{102,57,183}%
	\definecolor{aaltoTurquoise}{RGB}{0,168,180}%
	\definecolor{aaltoGreen}{RGB}{0,155,58}%
	\definecolor{aaltoLightGreen}{RGB}{105,190,40}%
	\definecolor{aaltoOrange}{RGB}{255,121,0}%
	\definecolor{aaltoFuchsia}{RGB}{177,5,157}%
\usepackage{soul}
\AddToHook{package/hyperref/after}{\def\htmladdnormallink#1#2{\href{#2}{#1}}}

\begin{document}

\renewcommand{\figurename}{FIG}
\renewcommand{\tablename}{TABLE}

\title{Effects of model size in density-functional-theory study of alloys: A case study of \texorpdfstring{$\text{CsPbBr}_2^{}\text{Cl}$}{}}

\author{Fang~Pan}
\thanks{These two authors contribute equally.}
\affiliation{State Key Laboratory for Manufacturing Systems Engineering; Electronic Materials Research Laboratory, Key Laboratory of the Ministry of Education, School of Electronic Science and Engineering, Xi'an Jiaotong University, Xi'an 710049, China \looseness=-2}
\author{Lin~Yang}
\thanks{These two authors contribute equally.}
\affiliation{State Key Laboratory for Manufacturing Systems Engineering; Electronic Materials Research Laboratory, Key Laboratory of the Ministry of Education, School of Electronic Science and Engineering, Xi'an Jiaotong University, Xi'an 710049, China \looseness=-2}
\author{Zhuangde~Jiang}
\affiliation{State Key Laboratory for Manufacturing Systems Engineering \& International Joint Laboratory for Micro/Nano Manufacturing and Measurement Technology, Xi'an Jiaotong University, Xi'an 710049, China \looseness=-2}
\author{Wei Ren}
\affiliation{State Key Laboratory for Manufacturing Systems Engineering; Electronic Materials Research Laboratory, Key Laboratory of the Ministry of Education, School of Electronic Science and Engineering, Xi'an Jiaotong University, Xi'an 710049, China \looseness=-2}
\author{Zuo-Guang~Ye}
\affiliation{Department of Chemistry and 4D LABS, Simon Fraser University, Burnaby, British Columbia V5A 1A6, Canada \looseness=-2}
\author{Jingrui~Li}
\email{jingrui.li@xjtu.edu.cn}
\affiliation{State Key Laboratory for Manufacturing Systems Engineering; Electronic Materials Research Laboratory, Key Laboratory of the Ministry of Education, School of Electronic Science and Engineering, Xi'an Jiaotong University, Xi'an 710049, China \looseness=-2}

\begin{abstract}
The primary challenge of density-functional-theory exploration of alloy systems concerns the size of computational model. Small alloy models can hardly exhibit the chemical disorder properly, while large models induce difficulty in sampling the alignments within the massive material space. We study this problem with the $\upgamma$ phase of the mixed halide inorganic perovskite alloy $\text{CsPbBr}_2^{}\text{Cl}$. The distribution of alloy formation energy becomes narrower when the size of the model system increases along $\,\sqrt[]{2}\times\,\sqrt[]{2}\times2$, $2\times2\times2$, and $2\,\sqrt[]{2}\times2\,\sqrt[]{2}\times2$ models. This is primarily because the distribution of $\text{Br}$ distribution parameters, which plays a leading role in determining the formation energy range, is more narrow for larger models. As a result, larger entropy stability effect can be observed with larger models especially at high temperatures, for which the approximation using mixing entropy based on the ideal solution model becomes better.
\end{abstract}


\maketitle
\thispagestyle{empty}



\section{Introduction}

Halide perovskites are promising candidates for next-generation optoelectronics owing to their outstanding electronic and optical properties, as well as the common starting materials and low synthesis cost \cite{Snaith13}. For example, the record power conversion efficiency of single-junction perovskite solar cells has broken the $26\%$ mark \cite{NRELchartALT,ZhouJ2024}, catching up with the conventional, expensive, and still market-dominating single-crystalline silicon devices \cite{MaF2023,Szabo2023}; the conversion efficiency against indoor white light has reached $\sim45\%$ \cite{MaQ2023,MaQ2024}; high-brightness, high external quantum efficiency, and excellent monochromaticity have been achieved with perovskite light-emitting diodes (PeLEDs) \cite{LuM2019,LiuXK2021,Fakharuddin2022}; and perovskite-based X-ray detectors recently exhibit high detection sensitivity, low detection limit, and low attenuation capability \cite{Yakunin2016,TsaiH2020,ZhangZ2024}. These emergent techniques are on their way to industrialization and commercialization, especially providing that the major barriers being overcome or circumvented, primary the instability of perovskite materials and devices against heat, moisture, and oxygen.

Alloying strategy is important and common in designing halide perovskite materials, as the many isostructural $\text{ABX}_3^{}$ ($\text{A}$, $\text{B}$, and $\text{X}$ denote the monovalent cation, bivalent cation, and halide anion, respectively) members offer a large flexibility in engineering the perovskite alloy composition toward desired materials properties \cite{CorreaBaena2017b,CorreaBaena2019,LuM2019}. Specifically, $\text{FA}_{1-x-y}^{}\text{MA}_x^{}\text{Cs}_y^{}\text{Pb}(\text{I}_{1-z}^{}\text{Br}_z^{})_3^{}$ ($\text{FA}=(\text{H}_2^{}\text{N})_2^{}\text{CH}^+$, formamidinium; and $\text{MA}=\text{CH}_3^{}\text{NH}_3^+$, methylammonium) with mixing at both $\text{A}$ and $\text{X}$ sites is one of today's most preferred core material for perovskite solar cells \cite{CorreaBaena2017,SunS2021}; $\text{B}$-site substitution (usually $\text{Sn}$ for $\text{Pb}$) is a common mean to reduce the environmental hazard of these emergent techniques because of lead; $\text{X}$-site alloying is the primary measure to tune the materials band gap, which is important for the desired emission wavelengths of PeLEDs \cite{Chiba2018,Karlsson2021}, and against different artificial light sources for indoor photovoltaics \cite{ZuoL2019,FengM2021,TongY2022}. Instability issues especially phase segregation are thus induced into the perovskite materials by alloying \cite{CorreaBaena2019,LuM2019,LiuL2021,LiuXK2021}.

First-principles calculations primarily using density functional theory (DFT) play an important role in studying the properties of perovskites including alloys. There are different approaches in previous studies for this purpose, mostly due to different sizes of model systems. Traverse DFT studies over all possible configurations (i.e., alignments of mixed anions) can be conducted for small model systems, such as the $8$-perovskite-unit model (i.e., $(\text{ABX}_3^{})_8^{}$) for $\text{FA}_{1-x}^{}\text{Cs}_x^{}\text{SnI}_3^{}$ ($8$ $\text{A}$ sites) \cite{GaoW2018} and the $4$-unit models for $\text{CsPb}(\text{I}_{1-x}^{}\text{Br}_x^{})_3^{}$ and $\text{CsPb}(\text{Br}_{1-x}^{}\text{Cl}_x^{})_3^{}$ ($12$ $\text{X}$ sites) \cite{PanF24}. Energy-minimizing algorithms such as Monte Carlo simulated annealing were employed to search for the minimal energy configurations of large model systems such as the $12$-unit model for $\text{FA}_{1-x}^{}\text{Cs}_x^{}\text{PbI}_3^{}$ ($12$ $\text{A}$ sites) \cite{YiC2016} and the $8$-unit model for $\text{CsPb}(\text{I}_{1-x}^{}\text{Br}_x^{})_3^{}$ and $\text{CsPb}(\text{Br}_{1-x}^{}\text{Cl}_x^{})_3^{}$ ($24$ $\text{A}$ sites) \cite{Laakso2022}. Alternatively, DFT calculations data of sampled configurations were used to train cluster-expansion models ($4$-unit models for binary mixed-halide perovskites \cite{Yin14b,Bechtel2018}) or machine-learning (ML) models ($8$-unit models for binary mixed-halide perovskites \cite{Laakso2022}). Notably, our group studied all four phases of both binary alloys $\text{CsPb}(\text{I}_{1-x}^{}\text{Br}_x^{})_3^{}$ and $\text{CsPb}(\text{Br}_{1-x}^{}\text{Cl}_x^{})_3^{}$ with traverse DFT calculations on the $4$-unit model systems (which are the smallest for both $\upgamma$ and $\updelta$ phases) \cite{PanF24}. For the first time, we have evaluated the thermodynamic state functions including Helmholtz free energy, internal energy, and configurational entropy for each phases and composition at different temperatures based on DFT data, with (i) correcting the conventional approach using on the mixing entropy based on the ideal solution model, and (ii) construct the temperature vs. composition phase diagrams for these alloys.

The biggest challenges for (DFT-based) computational study of alloys concerns the size and number of the model systems. In principle, we need large enough computational models that can properly exhibit the long-range order of alloys, whose short-range order is missing due to the ion or atom mixing (i.e., chemical disorder). And we need many configurations of alloyed ions to present the disorder character. This is, in principle, infeasible at least with pure DFT approach because of the massive computational resource demanded (note that the size of materials space grows with the combinatorial number related to the size of model system). Possible solutions should include: (i) to find the proper size of computational model system that can balance the computational cost and the reliability of results, and (ii) to develop advanced property-evaluating methods (such as ML based on DFT training data) whose accuracy is at the DFT level. To this end, size effects study of computational model systems for alloys is important. Note: The special quasirandom structures method \cite{Zunger1990,WeiSH1990} is widely used for modeling alloy materials. It uses Monte Carlo simulated annealing to minimize the difference between the cluster correlation functions of the constructed supercell configuration and those of the random structure. This method pays particular attention to the random alignment of alloyed ions or atoms. It also encounters the size-effects problem: the figure of merit of the best matching configuration decreases with the size of model system.

In this work, we study the inorganic mixed-halide perovskite alloy $\text{CsPbBr}_2^{}\text{Cl}$ in its room-temperature $\upgamma$ (space group $Pnma$) phase. This is a typical blue-light emitting material for PeLEDs \cite{WangC2020,KimYC2021} and closely related to candidate materials for harvesting ultraviolet light in indoor scenarios \cite{ZuoL2019,Chauhan2022}. Previous computational studies indicate that it is located at the formation free energy convex hull within a large temperature range \cite{PanF24}, and the fully regular configuration has the lowest alloy formation energy through the whole $\text{CsPb}(\text{Br}_{1-x}^{}\text{Cl}_x^{})_3^{}$ space \cite{Yin14b,Bechtel2018,Laakso2022,PanF24}. To study the effects of model size, we collect DFT calculation data from randomly sampled configurations of $2\times2\times2$ ($8$-unit) and $2\,\sqrt[]{2}\times2\,\sqrt[]{2}\times2$ ($16$-unit) model systems and compare them with the traverse DFT data of the smallest $\,\sqrt[]{2}\times\,\sqrt[]{2}\times2$ ($4$-unit) model. To estimate the impact of model size, we pay particular attention to the distribution of alloy formation energies, and eventually use this information to evaluate the thermodynamic state functions of each model.

The remainder of this article is organized as follows. In Sec.~\ref{models}, we briefly describe the investigated model systems and introduce some distribution parameters that distinguish the configurations. Section~\ref{results} presents the computational results, based on which a detailed discussion on the model size effects are conducted in Sec.~\ref{discussion}. Finally, Sec.~\ref{conclusion} concludes with a summary.

\section{Computational models of the \texorpdfstring{$\text{CsPbBr}_2^{}\text{Cl}$}{} alloy}\label{models}

\subsection{Structure models of the \texorpdfstring{$\text{CsPbBr}_2^{}\text{Cl}$}{} alloy}

\begin{figure}[ht]
{\centering
\small
\includegraphics[clip,trim=1in 7.9in 1in 1in,width=\textwidth]{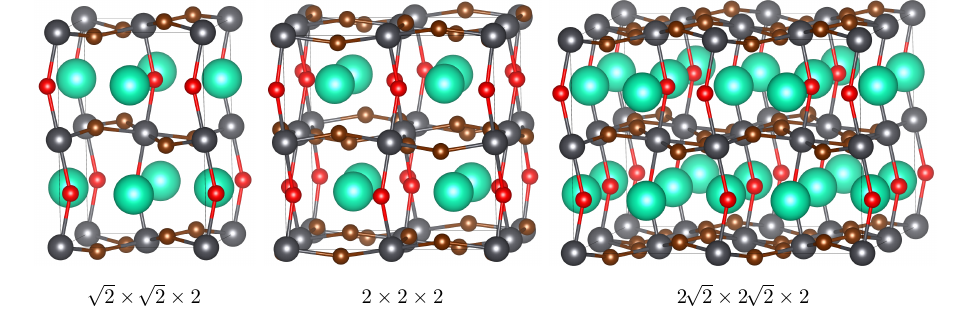}

}
\caption{Three structure models (from left to right: $\,\sqrt[]{2}\times\,\sqrt[]{2}\times2$, $2\times2\times2$, and $2\,\sqrt[]{2}\times2\,\sqrt[]{2}\times2$) for the $\upgamma$ phase of $\text{CsPbBr}_2^{}\text{Cl}$. $\text{Cs}$, $\text{Pb}$, in-plane halides, and out-of-plane halides are colored in green, dark gray, brown, and red, respectively.}
\label{fig:models}
\end{figure}

In this paper we focus on the alloy formation energy which is defined as
\begin{align}
E_i^{} &= E_{\text{total}}^{}(i) - \frac{2}{3}E_{\text{total}}^{}(\text{CsPbBr}_3^{}) - \frac{1}{3}E_{\text{total}}^{}(\text{CsPbCl}_3^{})
\end{align}
with $i$ indexing the $\text{CsPbBr}_2^{}\text{Cl}$ alloy configuration and $E_{\text{total}}^{}$ the DFT calculated total energy (per one perovskite unit, i.e., $5$ atoms). We considered three different models for the $\upgamma$ phase of $\text{CsPbBr}_2^{}\text{Cl}$, including the minimal one $\,\sqrt[]{2}\times\,\sqrt[]{2}\times2$ and two larger ones, $2\times2\times2$ and $2\,\sqrt[]{2}\times2\,\sqrt[]{2}\times2$ (Fig.~\ref{fig:models}). We must notice that in this context, the ``$\upgamma$ phase'' means that the configurations were constructed according to the following protocol: (i) a regular $\upgamma$ phase $\text{CsPbX}_3^{}$ structure was generated, which can be characterized by the $a^-b^-c^+$ Glazer tilting of the halide octahedra ($\text{X}_6^{}$); (ii) $\text{Br}^-$ and $\text{Cl}^-$ ions were randomly occupying the $\text{X}$ sites with the ratio $2:1$. In most cases, these structures do not exhibit the $Pnma$ symmetry, because (a) for the equivalent sites in $\text{CsPbX}_3^{}$, their local chemical environments are different in $\text{CsPbBr}_2^{}\text{Cl}$ because of the random $\text{Br}$/$\text{Cl}$ alignment, and as a result, (b) the octahedral tilting in the DFT relaxed structure is not as regular as in the pure $\text{CsPbBr}_3^{}$ or $\text{CsPbCl}_3^{}\,$. Nevertheless, they can still be categorized in the $\upgamma$ phase as the DFT relaxed structures exhibit $a^-b^-c^+$-like octahedral tilting.

The relaxed structure of each configuration and thus its properties are uniquely determined by the halide alignment. The first parameter is the $\text{Br}$ proportion in the in-plane (IP) halides. Here, in-plane means that the halides (brown in Fig.~\ref{fig:models}, labeled by $\bm{ab}$ hereafter) are approximately located in the plane spanned by the lattice vectors, around which the halide octahedral tilt out-of-phase. Out-of-plane (OP) halides (red in Fig.~\ref{fig:models}) are located between two neighboring $\bm{ab}$ planes. The possible numbers of IP $\text{Br}$ atoms and the corresponding numbers of possible alloy configurations of different models are listed in Table~\ref{tab:possibilities}. For example, the total number of $\text{Br}$/$\text{Cl}$ alignments with $6$ IP $\text{Br}$ in the $\,\sqrt[]{2}\times\,\sqrt[]{2}\times2$ is $C_8^6C_4^2=168$.

\begin{turnpage}
\begin{table}[ht]
\caption{IP and OP halide distributions in the three considered models. For each model, listed are possible numbers of IP $\text{Br}^-$ ions, the number of configurations at each IP $\text{Br}$ number, and its proportion in the total configurations.}
\label{tab:possibilities}
{\centering
\begin{tabular}{c@{\hspace{1em}}c@{\hspace{1em}}c@{\hspace{2em}}c@{\hspace{1em}}c@{\hspace{1em}}c@{\hspace{2em}}c@{\hspace{1em}}c@{\hspace{1em}}c} \hline\hline
\multicolumn{3}{c@{\hspace{2em}}}{$\,\sqrt[]{2}\times\,\sqrt[]{2}\times2$} & \multicolumn{3}{c@{\hspace{2em}}}{$2\times2\times2$} & \multicolumn{3}{c}{$2\,\sqrt[]{2}\times2\,\sqrt[]{2}\times2$} \\
Number of & Number of & & Number of & Number of & & Number of & Number of & \\
IP $\text{Br}^-$ ions & configurations & \multirow{-2}{*}{Proportion} & IP $\text{Br}^-$ ions & configurations & \multirow{-2}{*}{Proportion} & IP $\text{Br}^-$ ions & configurations & \multirow{-2}{*}{Proportion} \\ \hline
$4$ & ${\color{white}{0}}70$ & $0.1414$ 
& ${\color{white}{0}}8$ & ${\color{white}{0}}12870$ & $0.0175$
 & $16$ & $6.01\times10^{8{\color{white}{0}}}$ & $0.0003$ \\
$5$ & $224$ & $0.4525$ 
& ${\color{white}{0}}9$ & ${\color{white}{0}}91520$ & $0.1244$
 & $17$ & $9.05\times10^{9{\color{white}{0}}}$ & $0.0040$ \\
$6$ & $168$ & $0.3394$ 
& $10$ & $224224$ & $0.3049$
 & $18$ & $5.66\times10^{10}$ & $0.0251$ \\
$7$ & ${\color{white}{0}}32$ & $0.0646$ 
& $11$ & $244608$ & $0.3326$
 & $19$ & $1.95\times10^{11}$ & $0.0863$ \\
$8$ & ${\color{white}{00}}1$ & $0.0020$ 
& $12$ & $127400$ & $0.1732$
 & $20$ & $4.11\times10^{11}$ & $0.1822$ \\
& & & $13$ & ${\color{white}{0}}31360$ & $0.0426$
 & $21$ & $5.64\times10^{11}$ & $0.2499$ \\
& & & $14$ & ${\color{white}{00}}3360$ & $0.0046$
 & $22$ & $5.17\times10^{11}$ & $0.2291$ \\
& & & $15$ & ${\color{white}{000}}128$ & $0.0002$
 & $23$ & $3.21\times10^{11}$ & $0.1423$ \\
& & & $16$ & ${\color{white}{00000}}1$ & $1.36\times10^{-6}$
 & $24$ & $1.35\times10^{11}$ & $0.0600$ \\
& & & & & & $25$ & $3.85\times10^{10}$ & $0.0171$ \\
& & & & & & $26$ & $7.26\times10^{9{\color{white}{0}}}$ & $0.0032$ \\
& & & & & & $27$ & $8.80\times10^{8{\color{white}{0}}}$ & $0.0004$ \\
& & & & & & $28$ & $6.54\times10^{7{\color{white}{0}}}$ & $2.92\times10^{-5{\color{white}{0}}}$ \\
& & & & & & $29$ & $2.78\times10^{6{\color{white}{0}}}$ & $1.23\times10^{-6{\color{white}{0}}}$ \\
& & & & & & $30$ & ${\color{white}{00}}59520$ & $2.64\times10^{-8{\color{white}{0}}}$ \\
& & & & & & $31$ & ${\color{white}{0000}}512$ & $2.27\times10^{-10}$ \\
& & & & & & $32$ & ${\color{white}{000000}}1$ & $4.43\times10^{-13}$ \\
\multicolumn{9}{c}{In total} \\
& $495$ & $1$ & & $735471$ & $1$ & & $2.25\times10^{12}$ & $1$ \\ \hline\hline
\end{tabular}

}
\end{table}
\end{turnpage}

The total number of configurations are $C_{12}^8=495$, $C_{24}^{16}=735471$, and $C_{48}^{32}=2.25\times10^{12}$ for $\,\sqrt[]{2}\times\,\sqrt[]{2}\times2$, $2\times2\times2$, and $2\,\sqrt[]{2}\times2\,\sqrt[]{2}\times2$, respectively. For $\,\sqrt[]{2}\times\,\sqrt[]{2}\times2$, the number is reduced to $44$ because of symmetry thus allowing for a traverse DFT exploration \cite{PanF24}. While for the other two models, it is infeasible to perform DFT calculation on each configuration even symmetry can reduce the total number of configurations by some orders of magnitude. We sampled configurations of these two models according to the IP-OP distribution of halides as listed in Table~\ref{tab:possibilities}.


\subsection{Halide distribution parameters and machine learning}\label{ml-params}

The large number of mixed halide configurations in the large $2\times2\times2$ and $2\,\sqrt[]{2}\times2\,\sqrt[]{2}\times2$ structure models motivate to call of ML techniques. The aim is to rapidly predict the alloy formation energies of any given mixed halide alignments, to describe which we need a proper coding.

\begin{figure}[ht]
{\centering
\small
\includegraphics[height=6cm]{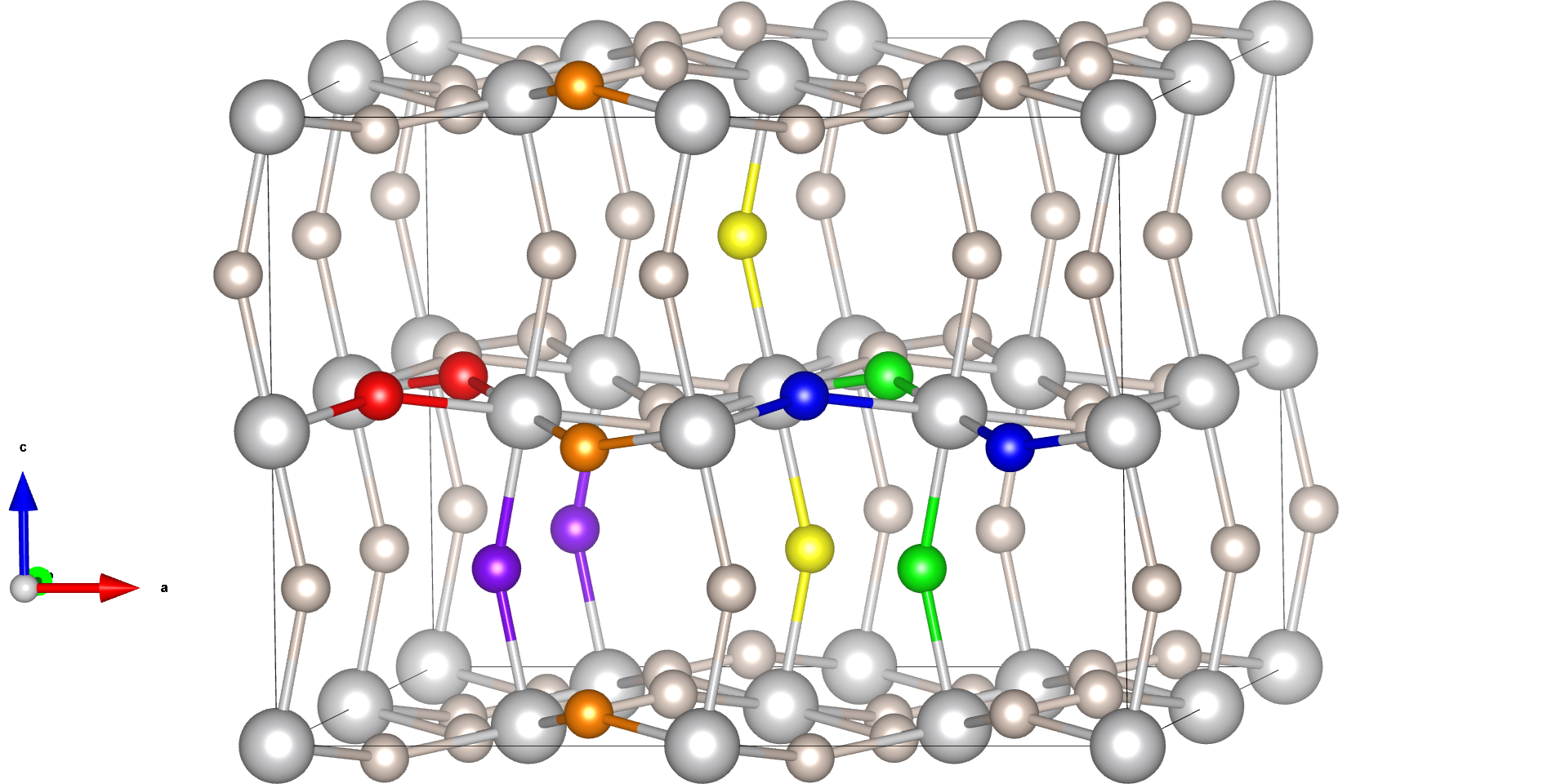}

}
\caption{$\text{Br}$\--$\text{Br}$ distribution parameters: SS, DS, OP, CP, SC, and CC (colored by red, blue, green, orange, yellow, and purple, respectively). Also shown is the coordinates $(\bm{a},\bm{b},\bm{c})$ system.}
\label{fig:param}
\end{figure}

We can use a set of parameters that describe the halide alignment. The IP/OP $\text{Br}$ proportion, denoted by $p_{\text{Br}}^{\text{IP}}$ hereafter, already introduced in Fig.~\ref{fig:models} and Table~\ref{tab:possibilities} provides the information of individual $\text{Br}^-$ ions' locations. In this regard it can be considered a one-body parameter. Two-body parameters give the information of the probability of two specific halides being found with a certain distance. These include
\begin{itemize}
  \item $p_{\text{SS}}^{}$: IP nearest-neighbor (NN) $\text{Br}$\--$\text{Br}$ proportion with both $\text{Br}^-$ ions on the same side (SS, i.e., above or below the plane in terms of $\bm{c}$) of the $\bm{ab}$ plane (Fig.~\ref{fig:param} red). The total number of SS-NN halide pairs of an $n$-unit model is $2n$.
  \item $p_{\text{DS}}^{}$: IP NN $\text{Br}$\--$\text{Br}$ proportion with two $\text{Br}^-$ ions on the different sides (DS) of the $\bm{ab}$ plane (Fig.~\ref{fig:param} blue). The total number of DS-NN halide pairs of an $n$-unit model is $2n$.
  \item $p_{\text{OP}}^{}$: OP NN $\text{Br}$\--$\text{Br}$ proportion with one IP and one OP $\text{Br}^-$ ions (Fig.~\ref{fig:param} green). The total number of OP-NN halide pairs of an $n$-unit model is $8n$.
  \item $p_{\text{CP}}^{}$: The portion of $\text{Br}$\--$\text{Br}$ pairs with both IP $\text{Br}^-$ ions at the neighboring $\bm{ab}$ planes (cross-plane or CP), with similar coordinates along both $\bm{a}$ and $\bm{b}$ (Fig.~\ref{fig:param} orange). The total number of OP-NN halide pairs of an $n$-unit model is $8n$.
  \item $p_{\text{SC}}^{}$: The portion of $\text{Br}$\--$\text{Br}$ pairs with both OP $\text{Br}^-$ ions at the same column (SC, i.e., with similar coordinates along both $\bm{a}$ and $\bm{b}$), connected with one $\text{Pb}^{2+}$ ion (Fig.~\ref{fig:param} yellow). The total number of SC halide pairs of an $n$-unit model is $n$.
  \item $p_{\text{CC}}^{}$: The portion of $\text{Br}$\--$\text{Br}$ pairs with both OP $\text{Br}^-$ ions at the nearest two columns (cross-column or CC), with similar coordinates along $\bm{c}$ (Fig.~\ref{fig:param} purple). The total number of SC halide pairs of an $n$-unit model is $2n$.
\end{itemize}
The whole set of parameters, $\{p_{\text{Br}}^{\text{IP}},p_{\text{SS}}^{},p_{\text{DS}}^{},p_{\text{OP}}^{},p_{\text{CP}}^{},p_{\text{SC}}^{},p_{\text{CC}}^{}\}$, are used as the first feature for our ML study.

Another feature for describing the atomic structure of $\text{CsPbBr}_2^{}\text{Cl}$ alloy is the many-body tensor representation (MBTR) \cite{HuoH2022}. MBTR is a structural descriptor that considers structural motifs such as elemental contents ($k=1$), interatomic distances ($k=2$), bond angles ($k=3$), etc. to form a vector representation of an atomic geometry \cite{HuoH2022,Himanen2020}. A previous study shows that the $k=1$ term does not improve the model accuracy when higher-order terms were included \cite{Stuke2019}, and it is the same for all configurations considered in this paper. It is generally concluded that the inclusion of the $k=3$ term produces a minimal improvement in accuracy with significantly increasing the computational cost \cite{Stuke2019,Laakso2022}. Therefore we only consider the $k=2$ term in this paper, similar as in previous studies \cite{Stuke2019,Laakso2022}. The second-order MBTR encodes the distances of all pairs of atoms of the investigated structure by
\begin{align}
\bm{M}_{\text{E}_1^{}\text{E}_2^{}}^{}(x) &= \sum_{l_1^{},l_2^{}} w_{l_1^{}l_2^{}}^{} \frac{1}{\varsigma\,\sqrt[]{2\uppi}} \exp\left( -\frac{\Big(x-R_{l_1^{}l_2^{}}^{-1}\Big)^2}{2\varsigma^2} \right)\,,
\end{align}
where $R_{l_1^{}l_2^{}}^{}$ is the interatomic (cartesian) distance between the $l_1^{}$th atom of element $\text{E}_1^{}$ and the $l_2^{}$ atom of $\text{E}_2^{}\,$, $\varsigma$ is the Gaussian width parameter. The sum runs over all atoms of element $\text{E}_1^{}$ (indexed by $l_1^{}$) and $\text{E}_2^{}$ (indexed by $l_2^{}$) within a cutoff distance, except that $l_1^{}\ne l_2^{}$ when $\text{E}_1^{}=\text{E}_2^{}\,$. The weighting function is defined by
\begin{align}
w_{l_1^{}l_2^{}}^{} &= \left\{
  \begin{array}{ll}
  \text{e}^{-sR_{l_1^{}l_2^{}}^{}}\,, & \quad\text{when}\quad R_{l_1^{}l_2^{}}^{}\leqslant r_{\text{cutoff}}^{}\,, \cr
  0\,, & \quad\text{when}\quad R_{l_1^{}l_2^{}}^{} > r_{\text{cutoff}}^{}\,,
  \end{array}
\right.
\end{align}
with $s$ controling its decay rate with $R_{l_1^{}l_2^{}}^{}\,$. In practice, one can limit all atoms of element $\text{E}_1^{}$ within the unit cell without loss of generality, and search for all atoms of $\text{E}_2^{}$ within the sphere of radius $r_{\text{cutoff}}^{}$ centered at this $\text{E}_1^{}$ atom, thus avoiding double counting.

The thus generated second-order MBTRs are continuous functions of $x$ whose dimension is the inverse of interatomic distance. They are then discretized at a set of equidistance grid points over a certain range of $x$ values. In this way, the MBTRs of all or particular atom pairs construct a vector presentation $\bm{M}$ of a given atomic structure.

MBTR has many advantages as a descriptor that maps the atomic structures to Hilbert space elements \cite{HuoH2022}. A particular merit is that MBTR, as a continuous function of the atomic coordinates (note: $R_{l_1^{}l_2^{}}^{}=\big\vert\left\vert\bm{R}_{l_2^{}}^{}-\bm{R}_{l_1^{}}^{}\right\vert$ with $\bm{R}_l^{}$ denoting the cartesian coordinate of the $l$th atom), allows a direct evaluation of its gradient. Therefore one can compute the atomic forces and stress tensor components directly from the ML model \cite{Laakso2022}. Nevertheless, we mainly focus on the relationship of how the formation energy of an alloy structure depends on the alignment of the alloyed element atoms in this paper. Thus we evaluate the MBTR of each configuration at its starting structure, which is initiate by linear interpolation between the relaxed $\text{CsPbBr}_3^{}$ and $\text{CsPbCl}_3^{}$ structures according to Vegard's law (i.e., the starting structures of different configurations only differ in the order of halide ions).

We correlate the ML features with the energy labels $E_{\text{label}}^{}$ using kernel-ridge regression (KRR) with the Gaussian kernel $k$ based on previous studies \cite{Stuke2019,Laakso2022} and tests:
\begin{align}
E_{\text{label}} &= \sum_i^N \beta_i^{}k(\bm{s},\bm{s}_i^{})\,, \\
k(\bm{s},\bm{s}_i^{}) &= \exp(-\gamma\vert\vert\bm{s}-\bm{s}^{\prime}\vert\vert_2^2)\,,
\end{align}
with $\bm{s}$ denoting the feature of a particular structure, $\beta_i^{}$ fitting coefficients, and $\gamma$ the width parameter controlling the kernel. The determination of $\{\beta_i^{}\}$ is done according to the following matrix form
\begin{align}
\bm{\beta} &= (\bm{K} + \alpha\bm{I})^{-1}\bm{E}_{\text{label}}^{\text{ref}}
\end{align}
with $\bm{K}$ the kernel matrix $K_{ij}^{}=k(\bm{s}_i^{},\bm{s}_j^{})$, $\bm{E}_{\text{label}}^{\text{ref}}$ the energy labels of the reference data set, and $\alpha$ the regularization parameter. The hyperparameters, including $\alpha$ and $\gamma$ and for the MBTR-ML case also $\sigma$, are optimized during the training of ML models using the Bayesian optimization structure search (BOSS) code \cite{Todorovic2019}.

\section{Results}\label{results}

\subsection{Benchmarking the DS-PAW calculations}

\begin{figure}[ht]
{\centering
\small
\includegraphics[height=8cm]{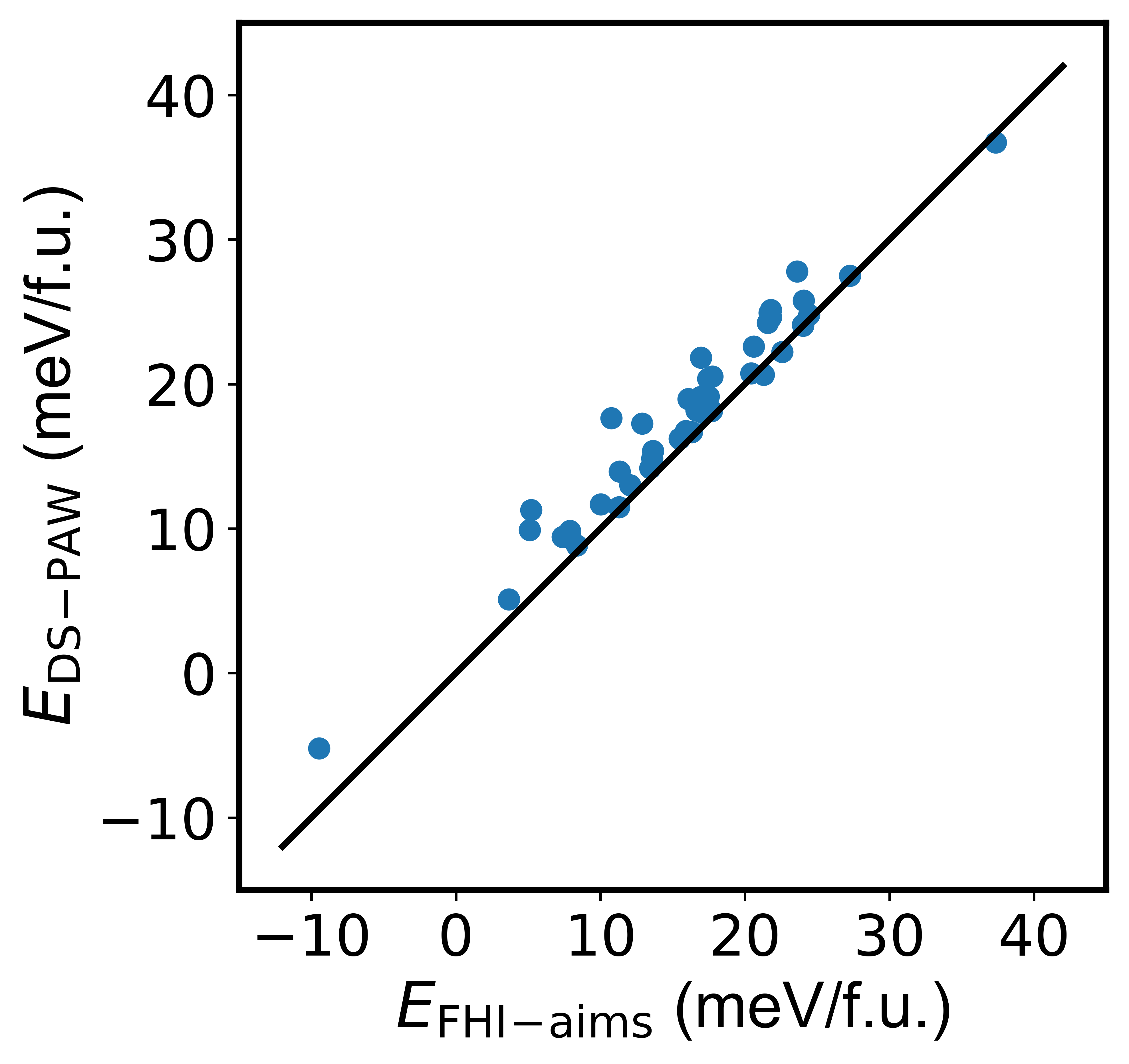}

}
\caption{Comparison between the alloy formation energies of all $44$ small-size models calculated by DS-PAW and FHI-aims.}
\label{fig:benchmark}
\end{figure}

Figure~\ref{fig:benchmark} compares the alloy formation energies of all $44$ small-size models of $\text{CsPbBr}_2^{}\text{Cl}$, obtained from structure optimization calculations using DS-PAW vs. FHI-aims (data from our previous study \cite{PanF24}). The results by these two softwares are generally similar, with the DS-PAW results slightly higher than the FHI-aims results. The root-mean-square error is $2.6~\text{meV}$ (per formula unit, similarly hereinafter), and the largest deviation is $6.9~\text{meV}$ Specifically, the portions of configurations with which the deviation is within $[0,1]~\text{meV}$, $(1,2]~\text{meV}$, and $(2,3]~\text{meV}$ are $\frac{17}{44}=0.39$, $\frac{10}{44}=0.23$, and $\frac{8}{44}=0.18$, meaning that the result difference between these two softwares are within $3~\text{meV}$ for $82\%$ of samples. Overall, linear refinement indicates that $\text{DS-PAW result}=0.94~\text{FHI-aims result}+3.0~\text{meV}$, with the coefficient of determination $R^2=0.95$.

The average expended resources for the DS-PAW and FHI-aims calculations are $56.6$ and $133.1~\text{CPU~h}$, respectively. Based on the comparison of computational accuracy and cost between these two softwares, we can conclude that DS-PAW performs well as it reaches the accuracy of FHI-aims which is widely used for halide perovskites \cite{LiJ16,LiJ18a,LiJ18b,Laakso2022,PanF24,LiJ2024} but requiring about only $40\%$ CPU hours.

\subsection{Energy distribution diagrams of model systems of different sizes}

Different to that we sampled all possible configurations of the smallest model system, the configurations of the medium and the large model systems were randomly sampled. Nevertheless, we expect that these random samples to contain most of the characters of the energy level distribution. To this end, we use density of states (DOS) to describe the distribution of alloy formation energies of the whole ensemble of $\text{CsPbBr}_2^{}\text{Cl}$ configurations for each model system size. It is defined by
\begin{align}
\rho(E) = A \sum_i g_i^{} \text{e}^{-(E-E_i^{})^2/\sigma^2}, & \quad
\int \rho(E) \operatorname{d}E = N
\end{align}
where $A$ is the normalization factor, $g_i^{}$ the degree of degeneracy and $E_i^{}$ the alloy formation energy of the $i$th configuration, $\sigma$ the width parameter (we chose $\sigma=0.5~\text{meV}$), $N$ is the total number of possible configurations. The smallest system is explored by a traverse DFT study, with $g_i^{}$ obtained from symmetry analysis (see Ref.~\cite{PanF24}). While for either the medium or the large system, the DOS is constructed based on $100$ randomly sampled configurations. As the sampling was performed according to a certain distribution of alloy configurations, we have all $g_i^{}=1$.

\begin{figure}[ht]
{\centering
\small
\includegraphics[height=8cm]{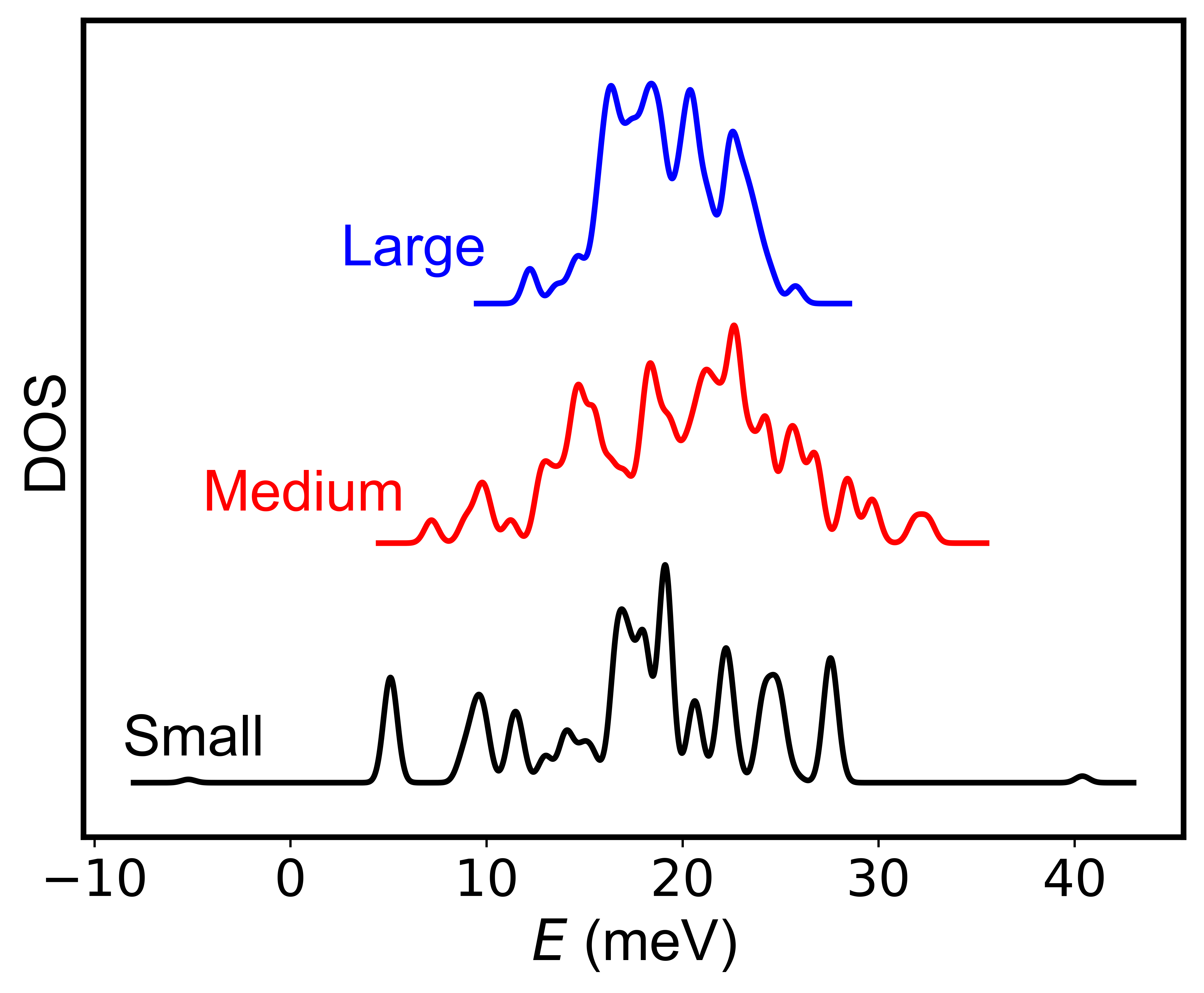}

}
\caption{Density of states (DOS) diagrams calculated with the smallest ($\,\sqrt[]{2}\times\,\sqrt[]{2}\times2$, in black), medium ($2\times2\times2$, in red), and large ($2\,\sqrt[]{2}\times2\,\sqrt[]{2}\times2$, in blue) models.}
\label{fig:dos}
\end{figure}

The DOS scheme (Fig.~\ref{fig:dos}) converts the discrete representation of (alloy formation) energy levels to a (quasi-)continuum of possibility distribution over a certain range of energies. It is obvious that as the model system size increases, the DOS distribution becomes narrower. Specific analysis is as follows. Firstly, the signals in the range $E<6~\text{meV}$ for the smallest model system vanish in the DOS for both larger model systems. The minimal-energy peak of the smallest model is at $-5.2~\text{meV}$, contributed from the fully ordered structure ($p_{\text{Br}}^{\text{IP}}=1$). The peak centered at $5.1~\text{meV}$ for the smallest model system is noticeable because of the high degree of degeneracy ($32$) of the corresponding configuration where $7$ $\text{Br}^-$ anions occupy the IP sites (i.e., $p_{\text{Br}}^{\text{IP}}=0.875$). However, configurations with both parameters were not sampled for both larger model systems due to the low possibilities (e.g., for $p_{\text{Br}}^{\text{IP}}=0.875$, the proportions are $4.6\times10^{-3}$ and $2.9\times10^{-5}$ for the medium and large models, respectively). Secondly, for the medium model system, configurations with alloy formation energy ranging between $28$ and $33~\text{meV}$ noticeably contribute to the DOS, while these configurations are not allowed by the smallest model system meaning that they are relatively largely disordered. Finally, both low- and high-energy ``tails'' of the medium model system DOS vanish for the large system, indicating that the corresponding configurations were not sampled for the latter. As a result, the arithmetic mean energy increases from the smallest ($18.0~\text{meV}$) to the medium ($20.1~\text{meV}$) model system, then slightly decreases when the model system size further increases ($19.2~\text{meV}$).

\subsection{Thermodynamics properties}

\begin{figure}
{\centering
\small
\includegraphics[clip,trim=1in 8in 1in 1in,width=\textwidth]{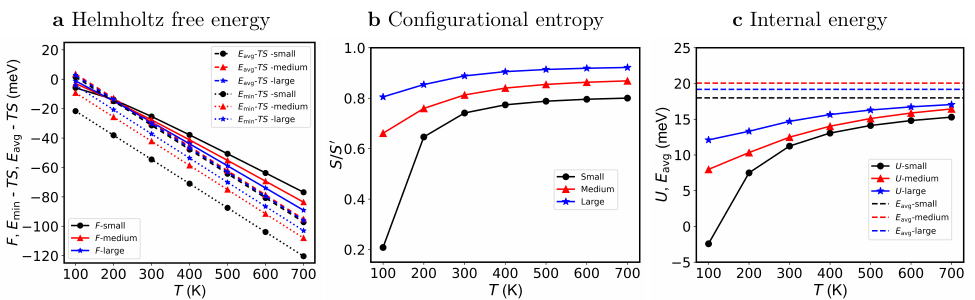}

}
\caption{Thermodynamic state functions of the smallest (black), medium (red), and large (blue) model systems calculated from their DOSs. \textbf{a} Helmholtz free energy. \textbf{b} configurational entropy in terms of its proportion to the value based on the ideal solution model. \textbf{c} internal energy. Also shown are: in \textbf{a}, Helmholtz free energies calculated by adding the mixing entropy term based on the ideal solution model to either the arithmetic mean energy (dashed lines) or the minimal formation energy (dotted lines); and in \textbf{c}, the arithmetic mean energy (dashed lines).}
\label{fig:thermo}
\end{figure}

From the DOS schemes we can calculate the partition function $Z$ and therewith the thermodynamic state functions, namely the Helmholtz free energy $F$, the internal energy $U$, and the configurational entropy $S$, for each model by
\begin{align}
Z &= \int \rho(E) \text{e}^{-nE/k_{\text{B}}^{}T} \operatorname{d}E\,, \label{partition} \\
F &= -\frac{1}{n}k_{\text{B}}^{}T \ln(Z)\,, \label{helmholtz} \\
U &= \frac{1}{nZ} \int E \rho(E) \text{e}^{-nE/k_{\text{B}}^{}T} \operatorname{d}E\,, \label{internal} \\
S &= \frac{U-F}{T} \label{entropy}
\end{align}
with $n$ denoting the number of perovskite units in the model system. The results are presented in Fig.~\ref{fig:thermo}.

The calculated Helmholtz free energy ($F$) lines (Fig.~\ref{fig:thermo}a) show a cross at $T=200~\text{K}$: $F(\text{small})<F(\text{medium})<F(\text{large})$ when $T<200~\text{K}$, while the opposite trend is observed when $T>200~\text{K}$. This can be rationalized by the alloy formation energy ranges of the sampled configurations: the contribution of the low-energy configurations of the smallest model system dominates the DOS at relatively low temperatures, while for the larger model systems, they were not sampled due to the low probabilities. As temperature increases, high-energy configurations play increasingly important role. The large ($2\,\sqrt[]{2}\times2\,\sqrt[]{2}\times2$) model exhibits the narrowest $\rho(E)$ distribution, thus showing the largest entropy stabilization effect as reflected by the smallest $F$ values.

When neglecting the volume change (which is adequate for solid states), we have $\operatorname{d}F=-S\operatorname{d}T$. At room temperature or higher ($T\geqslant300~\text{K}$), all three models exhibit an approximate linear dependence of $F$ on $T$ (Fig.~\ref{fig:thermo}a), indicating that the configurational entropy $S$ is nearly constant within this temperature range. Figure~\ref{fig:thermo}b clearly shows this character. Define the mixing entropy based on the ideal solution model by \cite{PanF24}
\begin{align}
S^{\prime} &= 3k_{\text{B}}^{} \bigg[ \frac{2}{3}\ln\bigg(\frac{2}{3}\bigg) + \frac{1}{3}\ln\bigg(\frac{1}{3}\bigg) \bigg]\,, \label{ideal}
\end{align}
Fig.~\ref{fig:thermo}b shows the evolution of $S/S^{\prime}$ with temperature. From $300$ to $700~\text{K}$, the proportions of configurational entropy in the ideal-solution-model-based mixing entropy of the smallest, medium, and large models increase from $74.1\%$ to $80.0\%$, from $81.3\%$ to $86.9\%$, and from $88.8\%$ to $92.2\%$, respectively. The configurational entropy trend stays at $S(\text{large})>S(\text{medium})>S(\text{small})$ within the whole temperature range, which is in accordance with the slope trend of $F$ of these three models shown in Fig.~\ref{fig:thermo}. As a final note, the configurational entropy of the smallest model is significantly smaller than the other two models at $100~\text{K}$, indicating that only a few configurations contribute to the thermodynamic properties of the whole ensemble of the smallest model at low temperatures.

Figure~\ref{fig:thermo}a provide an approximate method to evaluate the Helmholtz free energy of an alloy ensemble: subtracting the entropy term $TS^{\prime}$ obtained from the ideal solution model (Eq.~\ref{ideal}) from the arithmetic mean energy $E_{\text{avg}}^{}$ (dashed lines). From the definition of internal energy $U$ (Eq.~\ref{internal}), we know that $U$ it is always below $E_{\text{avg}}^{}$ unless all configurations have exactly the same energy value. $U$ approaches $E_{\text{avg}}^{}$ as temperature increases, as shown in Fig.~\ref{fig:thermo}c. As $E_{\text{avg}}^{}>U$ and $S^{\prime}>S$, $(E_{\text{avg}}^{}-TS^{\prime})$ might be a good approximation to $F=U-TS$ as both errors cancel each other to a certain extend.

Finally, we shortly discuss another way to evaluate the entropy stabilization effect which was widely used in computational studies \cite{YiC2016,GaoW2018,WangX2022}: subtracting the entropy term $TS^{\prime}$ obtained from the ideal solution model (Eq.~\ref{ideal}) from the (quasi)minimal alloy formation energy $E_{\min}^{}$, which can be obtained from DFT calculations on a series of test configurations and usually in combination with some minimization algorithm such as Monte Carlo simulated annealing \cite{Yin14b,Bechtel2018,Dalpian2019,Laakso2022}. According to Fig.~\ref{fig:thermo}a, this is a worse approach to $F$ compared to $(E_{\text{avg}}^{}-TS^{\prime})$. This is because $E_{\min}^{}<U$ and $S{\prime}>S$ therefore the errors of both terms lead to a smaller $F$ but can never cancel each other.

\section{Discussion}\label{discussion}

The results presented in Sec.~\ref{results} show noticeable size effects of the model systems of $\text{CsPbBr}_2^{}\text{Cl}$ alloy. They can be summarized as: the larger the model system, the narrower the distribution of alloy formation energy, the larger the configurational entropy (i.e., the closer to the mixing entropy based on the ideal solution model), and the smaller the Helmholtz free energy at room temperature or higher (i.e., the larger the entropy stabilization effect). In addition, large models are inappropriate to describe the low-temperature behavior of this alloy because the low-energy configurations were missed during the sampling.

\subsection{Energy distribution}

\begin{figure}[ht]
{\centering
\includegraphics[height=8cm]{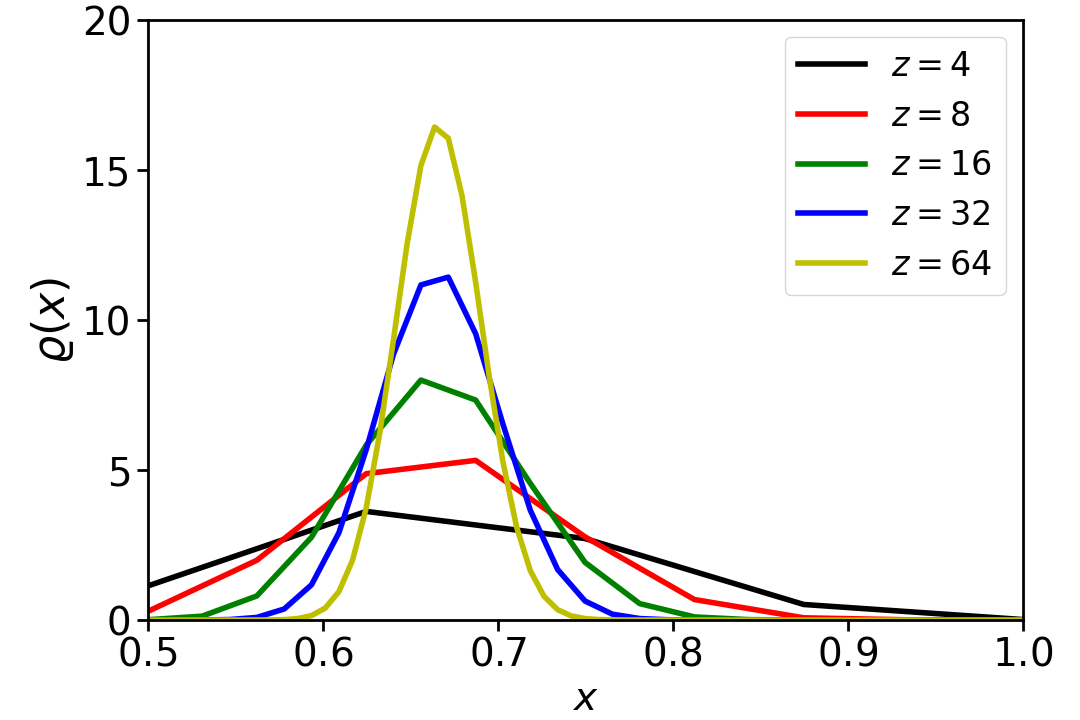}

}
\caption{Configuration distribution with respective to the IP $\text{Br}$ proportion parameter $p_{\text{Br}}^{\text{IP}}$ (labeled by $x$). Data of model systems of different sizes $z=4,8,16,32,64$ (only the first three considered in this paper) are colored in black, red, green, blue, and yellow, respectively.}
\label{fig:xdist}
\end{figure}

The proportion data listed in Table~\ref{tab:possibilities} can be converted into a quasi-continous distribution function $\varrho(x)$, with $x=\frac{m}{2n}=p_{\text{Br}}^{\text{IP}}$ denoting the $\text{Br}$ portion in all IP halides. Accordingly we have $\int\varrho(x)\operatorname{d}x=1$ with the integration running through the whole domain $x\in\Big[\frac{1}{2},1\Big]$. Figure~\ref{fig:xdist} shows the evolution of $\varrho(x)$ function with the model system size $z$ increases. From $\lim_{n\to+\infty}n!=\,\sqrt[]{2\uppi n}(n/\text{e})^n$, we have for the infinite model limit
\begin{align}
\lim_{n\to+\infty}\varrho(x) &= \lim_{n\to+\infty} 2n\frac{C_{2n}^mC_n^{2n-m}}{C_{3n}^{2n}} \nonumber \\
&\sim 
\exp\Big(-2n\Big[2(1-x)\ln(1-x)+x\ln(x)+\big(x-\textstyle\frac{1}{2}\big)\ln\Big(x-\textstyle\frac{1}{2}\Big)\Big]\Big) \label{x-distribution}
\end{align}
(more precisely, this is available within the major part of the domain $x\in\Big[\frac{1}{2},1\Big]$ except when $x$ is very close to the edge $\frac{1}{2}$ or $1$). Because of the coefficient $2n$ in the exponent, the exponential part dominates the character of $\varrho(x)$ with the following character: (i) $\varrho(x)$ exhibits a single, Gaussian-like peak at $x=\frac{2}{3}$, and (ii) the width of this peak decreases as $n$ increases. As a result, in the infinite model limit, almost all configurations will be characterized by $x=\frac{2}{3}$ which is natural because of the $\text{Br}_2^{}\text{Cl}$ stoichiometry.

\begin{figure}
{\centering
\small
\includegraphics[clip,trim=1in 8in 1in 1in,width=\textwidth]{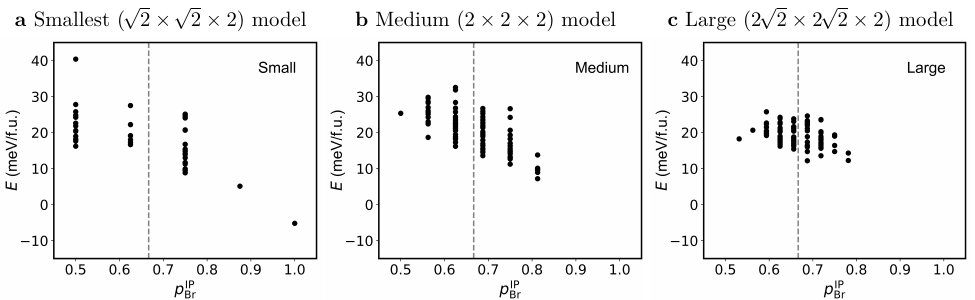}

}
\caption{Alloy formation energy distribution at each IP $\text{Br}$ proportion. \textbf{a} Smallest ($\,\sqrt[]{2}\times\,\sqrt[]{2}\times2$) model. \textbf{b} Medium ($2\times2\times2$) model. \textbf{c} Large ($2\,\sqrt[]{2}\times2\,\sqrt[]{2}\times2$) model. The vertical dashed lines mark $p_{\text{Br}}^{\text{IP}}=\frac{2}{3}\,$.}
\label{fig:e_br}
\end{figure}

Figure~\ref{fig:e_br} shows the alloy formation energies of sampled configurations at each IP $\text{Br}$ proportions. A general trend that the energy decreases with $p_{\text{Br}}^{\text{IP}}$ increases can be observed. The limit case of this trend was reported in previous studies: the minimal energy configuration exhibits a fully ordered $\text{Br}^-$/$\text{Cl}^-$ alignment so that all $\text{Br}^-$ anions occupy the IP halide sites \cite{Bechtel2018,Laakso2022,PanF24}; while in the maximal energy configuration, all OP halide sites are occupied by the larger halide $\text{Br}^-$, and accordingly the $\text{Br}^-$ proportion in IP halide sites is $0.5$ \cite{PanF24}.

Such a trend eases our analysis of the model-size dependence of the energy distribution by combining Figs.~\ref{fig:xdist} and \ref{fig:e_br}. The probability that the configurations with $p_{\text{Br}}^{\text{IP}}$ parameters close to $\frac{2}{3}$ are sampled increases with the size of model system. Fig.~\ref{fig:e_br} shows that the energies of these configurations are approximately distributed within the range $[10,30]~\text{meV}$, in good accordance with the arithmetic mean energies of all three models as alluded to. In addition, the energy distribution is narrower for larger model system, partly because larger model system allows configurations with $p_{\text{Br}}^{\text{IP}}$ closer to $\frac{2}{3}$ (the $p_{\text{Br}}^{\text{IP}}$ parameters closest to $\frac{2}{3}$ for the smallest, medium, and large model systems are $\frac{5}{8}$, $\frac{11}{16}$, and $\frac{21}{32}$, $0.042$, $0.021$, and $0.010$ away from $\frac{2}{3}$, respectively).

In principle, additional DFT calculation data are required to improve and eventually the DOS. However, this correspond to tremendous demand of computational resource and is infeasible for large model systems (such as the ``medium'' and ``large'' in this paper). Recently, the MBTR-KRR ML model has been successfully applied to inorganic mixed halide perovskites \cite{Laakso2022}. It can predict not only the formation energy but also the atomic forces of a $2\times2\times2$ supercell at the DFT accuracy ($\sim10^6$ and $\sim10^4$ faster, respectively), therefore it can perform structure optimization whose results are very similar to DFT. This would therefore be a proper choice if we want to construct the DOS of alloy formation energies in a (quasi)traverse manner. Nevertheless, it is natural to regard that the total energy of the relaxed structure of a particular $\text{CsPbBr}_2^{}\text{Cl}$ configuration uniquely depends on the ``topological'' $\text{Br}$/$\text{Cl}$ alignment, i.e., not including detailed atomic coordinate variation. We consider either the $\text{Br}$ distribution parameters defined in Sec.~\ref{ml-params} or only-halide (i.e., $\text{Br--Br}$, $\text{Cl--Cl}$, and $\text{Br--Cl}$) second-order MBTR satisfy this requirement.

\begin{figure}
{\centering
\small
\includegraphics[clip,trim=1in 6.6in 1in 1in,width=\textwidth]{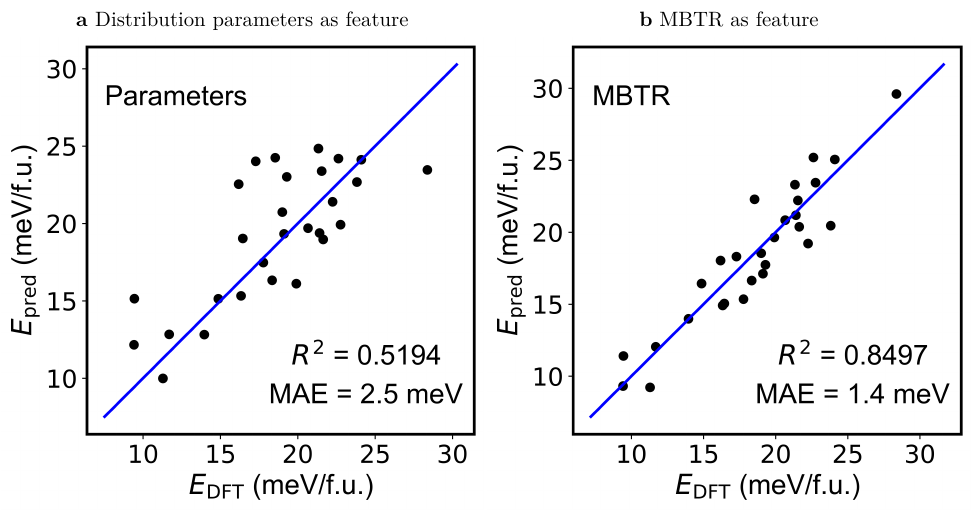}

}
\caption{KRR ML model of alloy formation energies of all configurations with $p_{\text{Br}}^{\text{IP}}\approx\frac{2}{3}\,$. \textbf{a} Using all $\text{Br}$ distribution parameters defined in Sec.~\ref{ml-params} as feature. \text{b} Using only-halide MBTR as feature.}
\label{fig:ml}
\end{figure}

Based on the DOS character of all three models, we only perform ML study on the configurations whose $p_{\text{Br}}^{\text{IP}}$ parameters are close to $\frac{2}{3}\,$. These include: $p_{\text{Br}}^{\text{IP}}=\frac{5}{8}$ and $\frac{6}{8}$ for the smallest ($\,\sqrt[]{2}\times\,\sqrt[]{2}\times2$) model, $p_{\text{Br}}^{\text{IP}}=\frac{10}{16}$ and $\frac{11}{16}$ for the medium ($2\times2\times2$) model, and $p_{\text{Br}}^{\text{IP}}=\frac{21}{32}$ and $\frac{22}{32}$ for the largest ($2\,\sqrt[]{2}\times2\,\sqrt[]{2}\times2$) model. The total number of data entries, $128$, is rather small because of the limited number of overall samples. We divided them as $100$ for training and $28$ for testing. Figure~\ref{fig:ml} shows the ML results using the $\text{Br}$-distribution parameters (Fig.~\ref{fig:ml}a) and halide-only MBTR (Fig.~\ref{fig:ml}b). Both ML models show good accuracy as reflected by the relatively small mean absolute error (MAE) values. $\text{Br}$-distribution-parameters-based ML model has a much lower $R^2$ coefficient than the MBTR-based model, in accordance with its larger MAE. Overall, the MBTR-KRR ML model performs better, and can fulfill the aim of approximately predicting the formation energy of the optimized structure of an alloy configuration.

\subsection{Thermodynamic properties}

From Eqs.~\ref{partition}\--\ref{entropy}, we can easily understand the observed trend of size effects on the thermodynamic properties at high temperatures. This is directly based on the interand when calculating $Z$ using Eq.~\ref{partition}, $\rho(E)\text{e}^{-nE/k_{\text{B}}^{}T}\,$. For larger $T$ and narrower $\rho(E)$ peak, the difference of $\text{e}^{-nE/k_{\text{B}}^{}T}$ values for energies within the peak is smaller. In the limit of $\frac{\langle\Delta E\rangle}{T}\to0$ ($\langle\Delta E\rangle$ denotes the energy deviation of the ensemble), we have
\begin{align}
Z &= \text{e}^{-n\langle E\rangle/k_{\text{B}}^{}T} \int \rho(E) \operatorname{d}E = \text{e}^{-n\langle E\rangle/k_{\text{B}}^{}T}\varOmega \,, \\
F &= \langle E\rangle-\frac{1}{n}k_{\text{B}}^{}T \ln(\varOmega)\,, \\
U &= \langle E\rangle, \\
S &= \frac{1}{n} \ln(\varOmega) = S^{\prime}
\end{align}
with $\varOmega$ denoting the number of configurations. The underlying physics is straightforward: since most of the configurations have nearly the same energy, their contribution to the thermodynamic properties of the whole ensemble are approximately equal. In this case, the ideal solution model is a good approximation.

\begin{figure}[ht]
{\centering
\includegraphics[height=8cm]{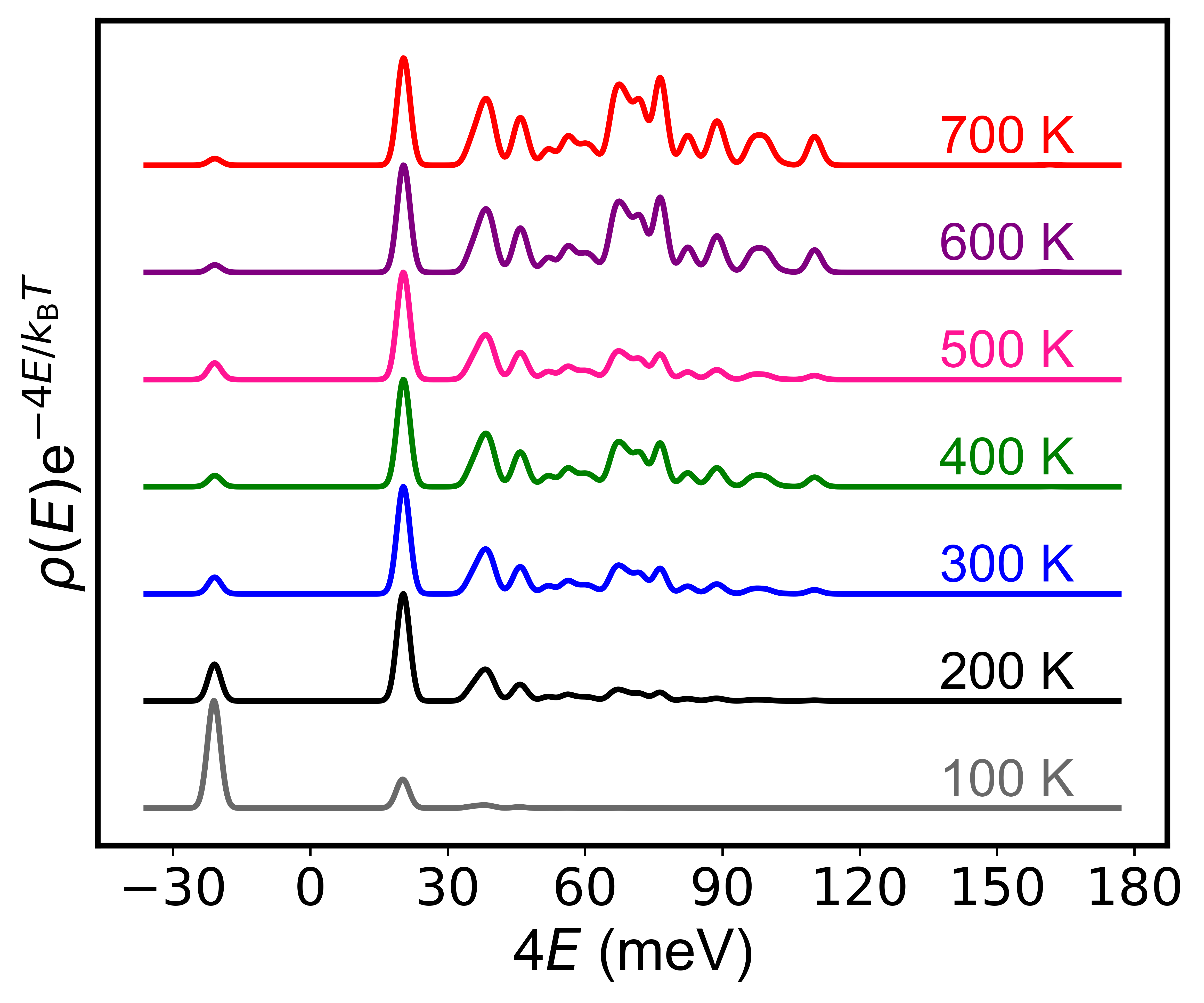}

}
\caption{Interand of partition function calculation (Eq.~\ref{partition}) of the smallest model system at different temperatures.}
\label{fig:z}
\end{figure}

The timeperature-evolution of the thermodynamic properties of the smallest model system can also be analyzed based on $\rho(E)\text{e}^{-nE/k_{\text{B}}^{}T}$ (Fig.~\ref{fig:z}). As the structural parameter distribution (Fig.~\ref{fig:xdist}) of this model system is rather flat, the $\rho(E)$ (DOS, Fig.~\ref{fig:dos}) is wide and structured. As a result, the $\rho(E)\text{e}^{-nE/k_{\text{B}}^{}T}$ function which exhibit the major contribution to the partition function $Z$ shows a transition from low- to high-energy regions as $T$ increases. When the temperature is as low as $100~\text{K}$, most of the contribution to the ensemble is provided by the minimal-energy configuration ($E=-5.2~\text{meV}$ or $4E=-20.8~\text{meV}$) despite the low degree of degeneracy ($1$). Accordingly, the configurational entropy is very small. At $200~\text{K}$, the contribution from this minimal-energy configuration rapidly decays, while the major contribution is from the configuration ($32$-fold degenerate, $E=5.1~\text{eV}$). As temperature further increases, configurations with higher energies becomes increasingly important, and accordingly the configurational entropy increases.

In sum, the high-temperature thermodynamic behavior of the inorganic halide perovskite alloy $\text{CsPbBr}_2^{}\text{Cl}$ can be properly evaluated, because the energy distribution of most of the configurations is relatively narrow and can already be adequately reflected by our large model that contains $16$ perovskite units. In contrast, the low-temperature behavior is more complicated. Minimal-energy configurations are naturally important at low temperatures. If their proportion is low (especially in large model systems), the evaluation of low-temperature thermodynamic properties of alloys remains open and will be subject to future work.

\section{Conclusion}\label{conclusion}

In summary, we explore the effects of model size on the computational study of thermodynamic properties of representative all-inorganic mixed-halide alloy $\text{CsPbBr}_2^{}\text{Cl}$. While the distribution of alloy formation energies for the smallest model is wide and structured, it becomes narrow when the size of model system increases, with configurations within both low- and high-energy regions being less sampled. As a result, the Helmholtz free energy of the alloy ensemble is more negative for larger model at room temperature or higher, indicating a stronger entropy stability effect. Overall, we observe obvious size effects of computational model systems for the $\text{CsPbBr}_2^{}\text{Cl}$, which rapidly decay when the model system becomes larger.

\section{Method}

For DFT structure optimization calculations, we chose the Perdew-Burke-Ernzerhof exchange-correlation functional for solids (PBEsol) \cite{Perdew2008} implemented in the DFT software DS-PAW, which is based on plane wave basis functions and projected augmented wave pseudo-potentials \cite{Bloechl1994} and distributed over the Device Studio platform \cite{DS}. PBEsol has already proved its suitability in modeling halide perovskites \cite{YangRX2017,Bokdam2017,Seidu2021a,Laakso2022,LiJ2024} and was used for the calculations of $\text{CsPbBr}_2^{}\text{Cl}$ alloys \cite{PanF24}, whose data were used as reference in this paper. The plane-wave cutoff was set to $600~\text{eV}$ based on test calculations. Structural relaxation was performed using the conjugate gradient method with the convergence criteria for total energy and atomic forces set to $10^{-4}~\text{eV}$ and $0.05~\text{eV}\cdot\text{\AA}^{-1}$, respectively. $\Gamma$-centered $8\times8\times6$, $6\times6\times6$, and $4\times4\times6$ $k$-point meshes were used for the Brillouin zone integration for the $\,\sqrt[]{2}\times\,\sqrt[]{2}\times2$, $2\times2\times2$, and $2\,\sqrt[]{2}\times2\,\sqrt[]{2}\times2$ model systems, respectively.

\section{Acknowledgement}

The authors thank Hua Dong, Jie Xu, Fang Yuan, and Jarno Laakso for fruitful discussions. We gratefully acknowledge HZWTECH for providing computational facilities. We also thank the computational resources provided by Xi'an Jiaotong University's HPC platform and the Computing Center in Xi'an. F.P. thanks Shenbo Yang and Duo Zhang from HZETECH for technical assistance. This work was supported by the Natural Science Foundation of Shaanxi Province of China (grant No. 2023-YBGY-447) and the National Natural Science Foundation of China (grant No. 62281330043).


\end{document}